\documentclass{nature}

\usepackage{color}
\usepackage{graphicx}
\usepackage{verbatim}
\makeatletter
\let\saved@includegraphics\includegraphics
\AtBeginDocument{\let\includegraphics\saved@includegraphics}
\renewenvironment*{figure}{\@float{figure}}{\end@float}
\makeatother

\title{Faster, farther, stronger: spin transfer torque driven high-order propagating spin waves in nano-contact magnetic tunnel junctions}

\author{A. Houshang$^{1,2}$, R. Khymyn$^{1}$, M. Dvornik $^{1}$, M. Haidar$^{1}$, S. R. Etesami$^{1}$, R. Ferreira$^{3}$, P. P. Freitas$^{3}$, R. K. Dumas$^{1}$ \& J. \AA{}kerman$^{1,2,4}$}

\begin{document}

\maketitle

\begin{abstract}
Short wave-length exchange-dominated propagating spin waves will enable magnonic devices to operate at higher frequencies and higher data transmission rates.\cite{Yu2016} While GMR based magnetic nano-contacts are highly efficient injectors of propagating spin waves\cite{madami2011nn,Urazhdin2014nn}, the generated wave lengths are 2.6 times the nano-contact diameter\cite{Slonzewski1999jmmm}, and the electrical signal strength remains much too weak for practical applications. Here we demonstrate 
nano-contact based spin wave generation in magnetic tunnel junction stacks, and observe large discrete frequency steps consistent with the hitherto ignored possibility of second and third order propagating spin waves with wave lengths of 120 and 74 nm, i.e.~much smaller than the 150 nm nano-contact.
These higher-order propagating spin waves will not only enable magnonic devices to operate at much higher frequencies, but also greatly increase their transmission rates and spin wave propagating lengths, both proportional to the much higher group velocity. 
\end{abstract}

Steady state large angle magnetization dynamics can be generated via spin transfer torque (STT)\cite{Berger1996,Slonczewski1996,Slonzewski1999jmmm} in a class of devices commonly referred to as spin torque nano-oscillators (STNOs)\cite{Tsoi1998,Myers1999,Chen2016procieee} 
The typical building block of an STNO is a thin film trilayer stack, where two magnetic layers are separated by a nonmagnetic spacer. The charge current becomes partially spin polarized by the magnetic layers and can act as positive or negative spin wave (SW) damping, depending on its polarity. 
Above a certain critical current density, the negative damping can locally overcome the intrinsic damping resulting in auto-oscillations on one or more SW modes of the system. To sustain such auto-oscillations, a large current density of the order of $10^{6}-10^{8}$ A/cm$^2$ is required, which can be achieved by spatial constriction of the current, \emph{e.g.} using a nano-contact (NC) on top of a GMR trilayer stack. Such NC based STNOs are also the most effective SW injectors for miniaturized magnonic devices\cite{madami2011nn,Urazhdin2014nn}, in particular for short wave length, exchange-dominated SWs, since the wave vector ($k$) is inversely proportional to the NC radius ($r_{NC}$) through the Slonczewski relation $k=1.2/r_{NC}$. As the SW group velocity, which governs the data transmission rate, scales with $k$, and the operating frequency with $k^2$, future ultra-high data rate magnonic devices will have to push the SW wave length down to a few 10s of nanometers.\cite{Yu2016} For efficient electrical SW read-out, magnonic devices will also have to be based on magnetic tunnel junctions (MTJs), as tunneling magnetoresistance (TMR) is one or more orders of magnitude higher than GMR.\cite{Yuasa2004,Parkin2004} 

\begin{figure}
  \begin{center}
  \includegraphics[width=12 cm]{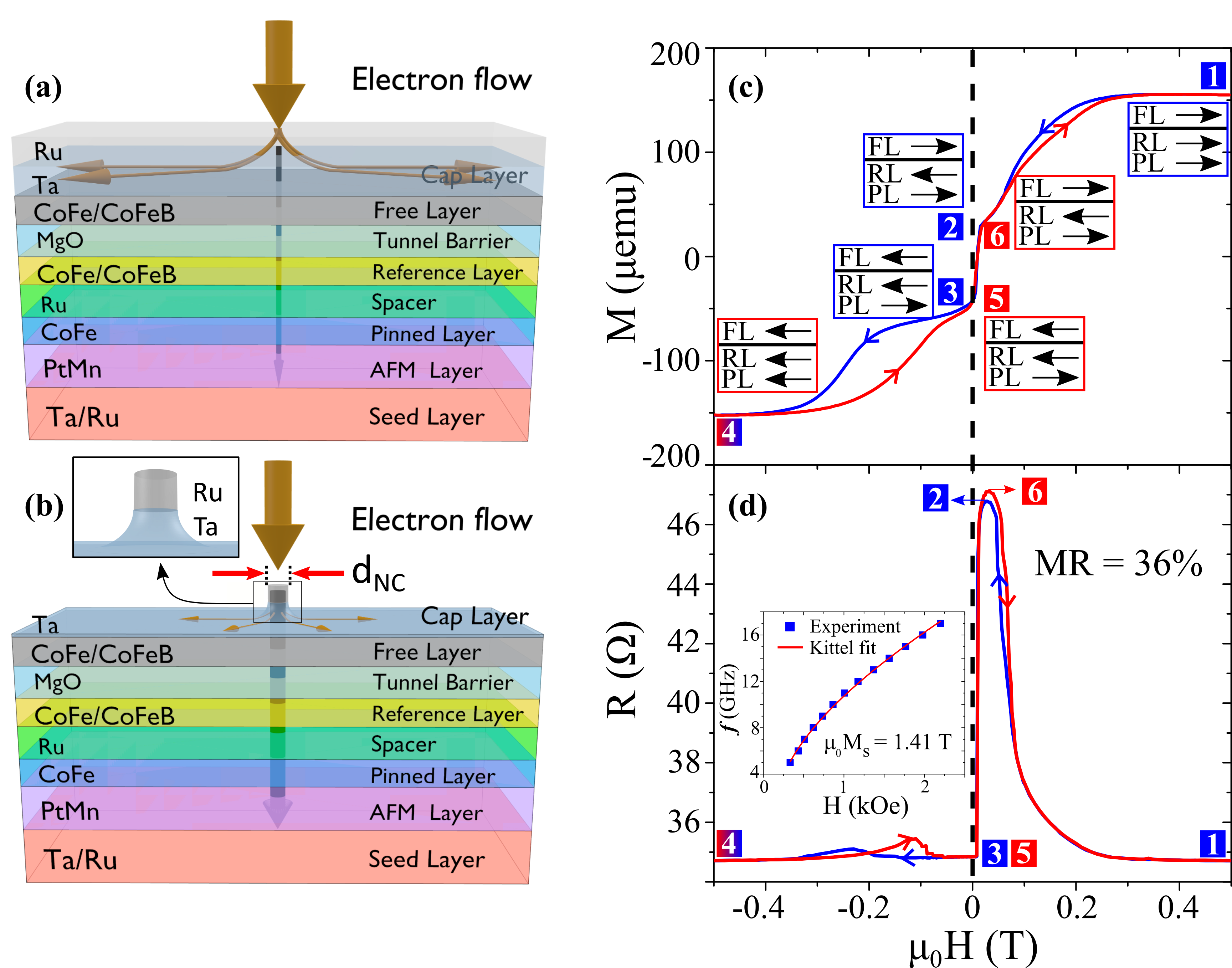}
    \caption{\textbf{Device schematic, current distribution, and static measurements.} (a) Schematic of the material stack showing the current distribution (a) for an ordinary NC, and (b) a NC where the Ta/Ru cap has been ion milled into a pillar (inset shows the remaining Ta/Ru) .
    The ion milled pillar reduces the shunt current (orange arrows) in the cap layer and forces a larger fraction of the current to go through the stack (black arrow). (c) Hysteresis loop of the MTJ stack before patterning, with the magnetic field applied along the in-plane easy axis. The magnetic state of the three magnetic layers (free (FL), reference (RL), and pinned layer(PL)) is depicted by the three arrows at six points along the hysteresis loop.
    (d) The resistance of the final device measured vs.~magnetic field along the easy axis showing MR of 36\%. Inset in (d) is the frequency of the uniform FMR mode of the free layer as a function of in-plane field. Red solid line is a Kittel fit to extract an effective magnetization of 1.41 T.
    }
    \label{fig-device}
    \end{center}
\end{figure}

The relatively low conductivity of the MTJ tunnelling barrier compared to the top metal layers leads to large lateral current shunting for an ordinary NC (Fig.1(a)). To force more of the current through the MTJ, we instead fabricated so-called ``sombrero" NCs (Fig.~1(b)), in which the MTJ cap layer is gradually thinned as it extends away from the NC.\cite{Maehara2013,Maehara2014} This ensures that a larger fraction of the current passes \emph{through} the MTJ while simultaneously keeping the free layer intact.
Using a MgO layer with a low resistance-area (RA) product (here 1.5 $\Omega \ \mu \mathrm{m}^2$) further promotes tunnelling through the barrier.
Fig.~1(c) shows the magnetic hysteresis loop of the unpatterned MTJ stack in a magnetic field applied along the in-plane easy axis (EA) of the magnetic layers (for details see Methods). Fig.~1(d) shows the corresponding resistance ($R$) of an MTJ-STNO with a nominal diameter of $d_{NC}$\,=\,150 nm, displaying a magnetoresistance of 36$\%$, confirming that a significant fraction of the current indeed tunnels through the MgO barrier. 
The very good agreement between the field dependence of the unpatterned stack and the fully processed MTJ-STNO suggests minimal process-induced changes of the magnetic layers, a strong indication that the free layer remains intact.

\begin{figure}
  \begin{center}
  \includegraphics[width=16cm]{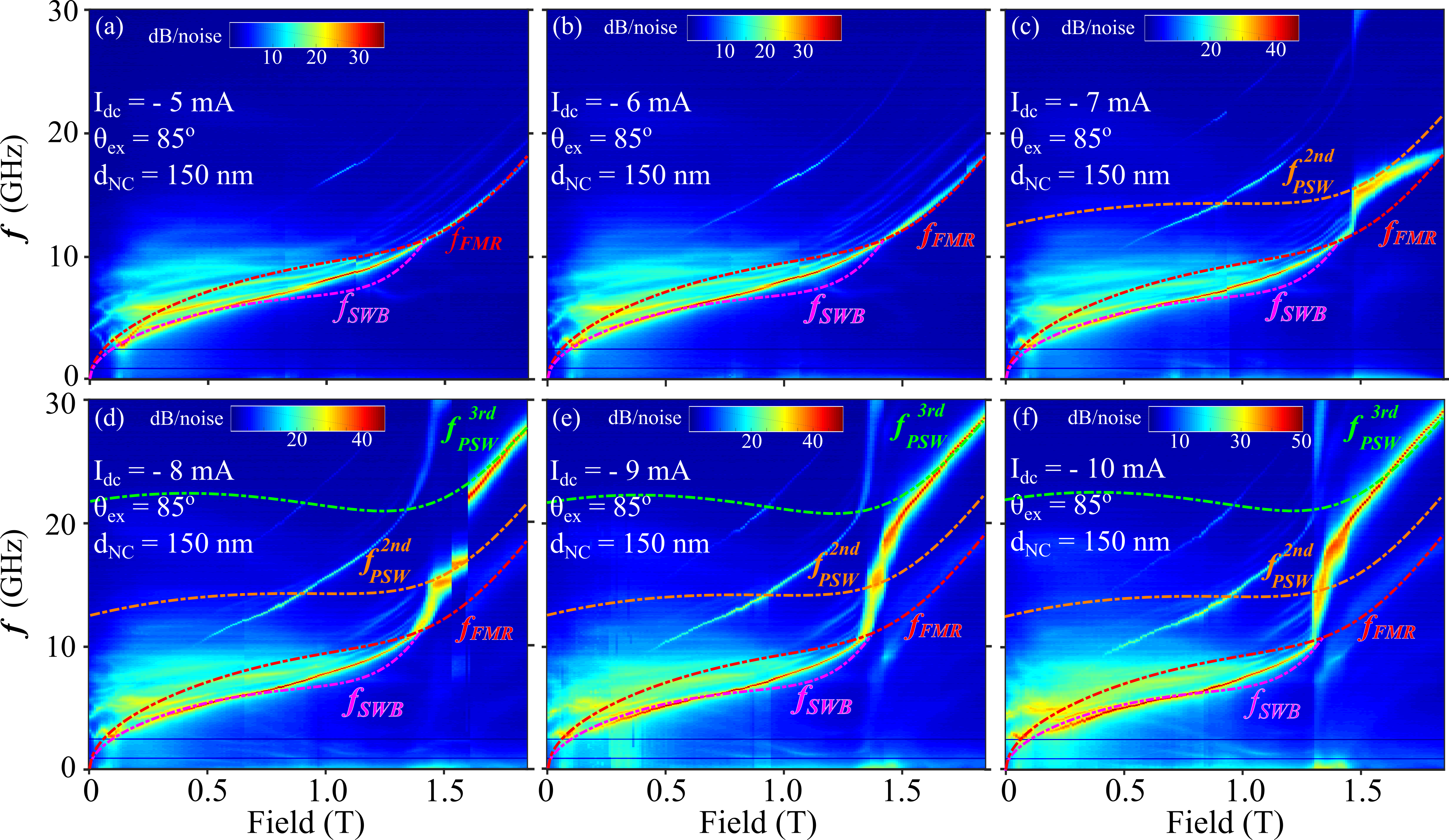}
    \caption{\textbf{Auto-oscillations vs.~field strength: Higher order Slonczewski modes.} (a)-(f) PSD vs.~applied field strength ($\theta_{ex} = 85^{\circ}$) and for currents (a) $I_{dc} = -5$ mA, (b) $\mathrm{I}_{dc} = -6$ mA, (c) $I_{dc} = -7$ mA, (d) $I_{dc} = -8$ mA, (e) $I_{dc} = -9$ mA, and (f) $I_{dc} = -10$ mA.  The pink, red, brown, and green dashed lines represent the calculated frequencies for the SW bullet ($\mathit{f_{SWB}}$), the FMR ($\mathit{f_{FMR}}$), the 2$^{nd}$ order Slonczewski mode ($\mathit {f_{PSW}^{2nd}}$), and the 3$^{rd}$ order Slonczewski mode ($\mathit {f_{PSW}^{3rd}}$), respectively.}
    \label{fig2}
    \end{center}
\end{figure}

Fig. 2 shows the generated power spectral density (PSD) vs.~field strength during auto-oscillations at six different drive currents, with the field angle fixed to $\theta_{ex}$ = 85$^\circ$. At the lowest currents, $I_{dc} = -5 \mathbin{\&} -6$ mA,
the strongest mode can be identified as a SW bullet soliton\cite{slavin2005prl,bonetti2010prl,dumas2013prl}. Its frequency, $f_{SWB}$ lies well below the ferromagnetic resonance frequency ($f_{FMR}$; red dashed line) and can be very well fitted (Eq.~4 in Methods) for fields below 0.7 T. 
At intermediate fields, 0.7 T $<H_{ex}<$ 1.35 T, the bullet signal gradually weakens and its frequency 
approaches $f_{FMR}$, until at 1.35 T it finally disappears as its frequency crosses $f_{FMR}$ where the self-localization of the bullet is no longer possible.  
The calculated (Eq. (\ref{eq-internal}) in Methods) internal angle of the magnetization, $\theta_{int}^{crit}=60^\circ$, at the critical field $\mu_0 H_{ex}=1.35$ T is in good agreement with the theoretical prediction\cite{gerhart2007prb} $\theta_{int}^{crit}=55^\circ$. Above the critical field, we find a weaker mode about 0.2 GHz above $f_{FMR}$ consistent with the ordinary Slonczewski propagating SW mode\cite{Slonzewski1999jmmm}.

At higher currents, $I_{dc}=-7$ mA, the PSD in the high-field region changes dramatically, as a much stronger mode appears with a frequency much higher than $f_{FMR}$. This change is accompanied by additional low-frequency noise, indicative of mode hopping. Further increasing the current to $I_{dc}=-8$ mA, first modifies this new mode, after which another sharp jump up to an even higher frequency is observed at about 1.6 T. As we increase the current magnitude further to -9 \& -10 mA, this new mode dominates the entire high-field region. The low-frequency noise is now concentrated to the field region just above the critical field, where the 2$^{nd}$ and 3$^{rd}$ modes appear to be competing. 

To analyze this behavior, we draw renewed attention to the higher-order propagating SW modes put forward by Slonczewski
\cite{Slonzewski1999jmmm} but up to this point entirely overlooked in experiments. In the linear regime, the excited SWs have a discrete set of possible wave vectors $r_{NC} k \simeq 1.2, 4.7, 7.7 ...$, where only the first order mode ($r_{NC} k \simeq 1.2$) is discussed in the literature, because of its lower threshold current. Taking the literature value\cite{devolder2011spin} for the free layer exchange stiffness, $A_{ex} = 23\times10^{-12}$ J/m, and allowing for a reasonable lateral current spread (an effective NC radius of $r_{NC}=90$ nm), we find that we can fit the field dependent frequencies of both the 2$^{nd}$ and 3$^{rd}$ mode almost perfectly, using the predicted $k=4.7/r_{NC}$ and $k=7.7/r_{NC}$. The ordinary first-order mode can be equally well fitted (not shown). It is noteworthy that increasingly higher currents are required to excite the higher mode numbers, in agreement with Slonczewski's original expectations\cite{Slonzewski1999jmmm}. We also point out that in the original derivation only radial modes were considered, excluding any azimuthal modes. The additional smaller frequency step observed within the 2$^{nd}$ radial mode in Fig.~2(d) could hence be related to a further increase in the exchange energy of a possible azimuthal mode.\cite{Arias2007}

In all fits, we allowed $M_{s}$ to be a function of $I_{dc}$ and used the same $M_{s}$ to calculate $f_{FMR}$, $f_{SWR}$, and $f_{PSW}^{i=1,2,3}$. 
This allows us to estimate the amount of heating due to the drive current. Fig.~3(a) shows the variation of $M_{s}$ as a function of temperature (blue circles) measured from 10 to 340 K using temperature dependent FMR spectroscopy on unpatterned areas of the MTJ stack. We can fit $M_{s}(T)$ to a Bloch-law function and extrapolate this dependence to higher temperatures (black solid line). The red triangles in Fig.~3(a) then shows the extracted $M_{s}$ values at each $I_{dc}$, placed on the extrapolated fit, which allow us to extract the local temperature of the free layer underneath the NC.
As can be seen in the inset, the temperature shows a parabolic rise with increasing current, indicative of Joule heating. The current-induced temperature rise at e.g. $I_{dc} = -9$ mA is about 220 K, which is consistent with literature values of nanoscale temperature gradients in similar structures sustaining similar current densities\cite{Petit2012prl}.

\begin{figure}
  \begin{center}
  \includegraphics[width=12cm]{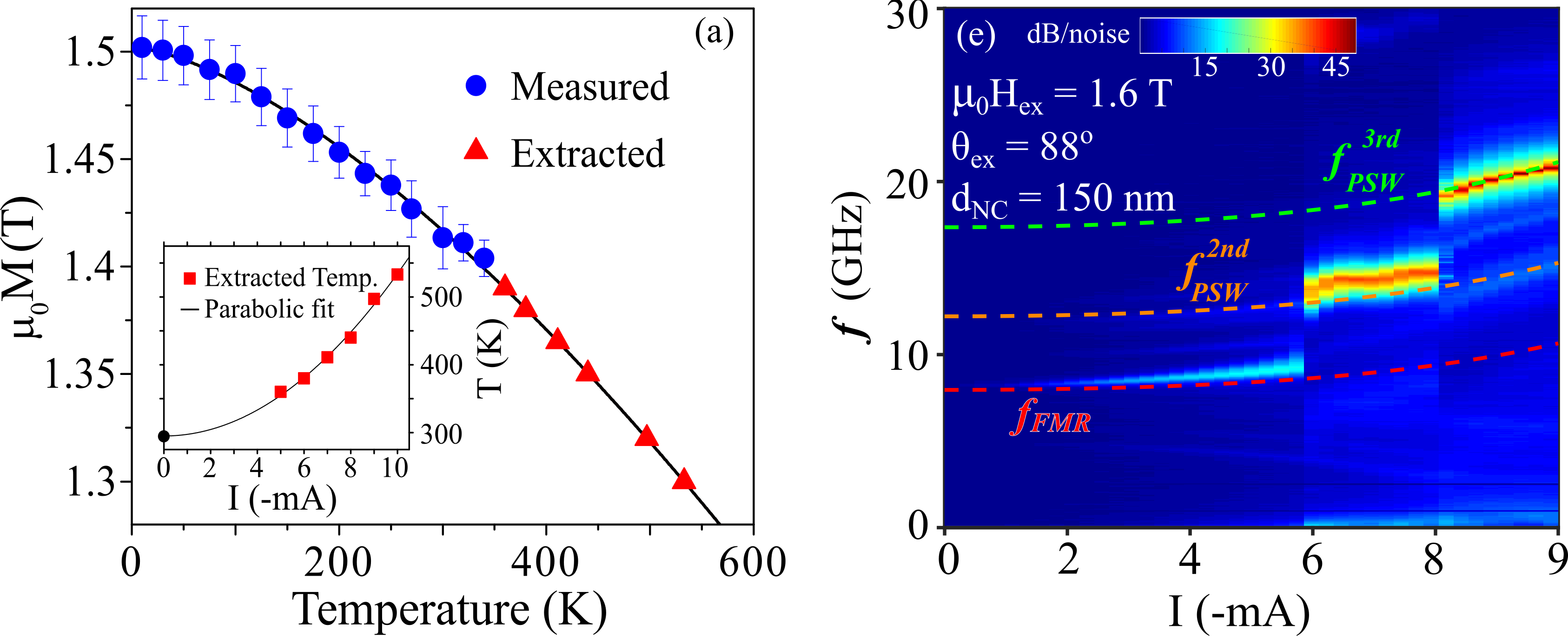}
    \caption{\textbf{Temperature dependence of the magnetization, and current control of the 
    higher-order Slonczewski modes.}
    (a) Temperature dependence of $M_S$ where the blue data points are CryoFMR measurements on unpatterned areas of the MTJ wafer, the black solid line is a Bloch law fit to these data points, and the red triangles are the extracted values for $M_S$ from fits to the data in figure 2(a)-(f) placed on the extrapolated part of the Bloch law. The corresponding temperatures are plotted vs.~STNO current in the inset together with a fitted parabola, fixed at room temperature for zero current. (b) PSD vs.~drive current for another sample with the same nominal size at $\theta_{ex} = 88^{\circ}$ and $\mu_{0} H_{ex}$ = 1.6 T. }
    \label{fig3}
    \end{center}
\end{figure}

We finally show how we can control which propagating mode to excite by varying the current at constant applied field (Fig.~3(b)). We can again fit the three modes very accurately, using the current dependent $M_{s}(I_{dc})$ extracted from Fig.~3(a). 
The weak current tunability of our MTJ based NC-STNOs is consistent with the weak non-linearity values found in the literature on MTJ pillars\cite{Kudo2009apl,Bianchini2010apl} and is advantageous as it reduces any non-linearity driven increase in phase noise from amplitude noise\cite{Slavin2009,Lee2013prb}.

The possibility of generating higher order Slonczewski modes has a number of important implications. Their much shorter wavelength, in our case estimated to 120 nm (2$^{nd}$ mode) and 74 nm (3$^{rd}$ mode), already bring them into the important sub-100 nm range\cite{Yu2016}, which only a few years ago was considered out-of-reach for magnonics\cite{chumak2015}. As the SW group velocity increases linearly with the wave vector as $v_{gr} \simeq 4 \gamma A_{ex} k/M_s $, much faster transmissions can be achieved in magnonic devices and the SWs can travel significantly farther before being damped out. The calculated group velocities for the three observed modes are $v_1=258$ m/s, $v_2=1010$ m/s and $v_3=1655$ m/s.   This will be particularly beneficial for mutual synchronization of multiple MTJ based nano-contacts\cite{mancoff2005nt,kaka2005nt,sani2013ntc,Houshang2016ntn}. 
For example, one can find the maximum distance of synchronization $a_{max}$ between two coupled oscillators, using the method developed by Slavin and Tiberkevich\cite{Slavin2006theory}. Using typical parameters of coupling strength, $\Delta_{max}/\sqrt{1+\nu^2}=50$~MHz, and a Gilbert damping of $\alpha_G=0.015$, we find $a_{max}=240, 350, 420$~nm for the PSWs with the corresponding frequencies $f_{PSW}=13.5, 17.7, 24.7$~GHz observed at $\mu_0 H_{ext}=1.6$ T (Fig. \ref{fig3}). As the drive current can be varied very rapidly, high data rate frequency shift keying will also be possible using only a small modulation amplitude of the drive current. In addition, novel modulation concepts such as wave vector keying could be readily realized, with possible use in magnonic devices.

We conclude by pointing out that both the nominal NC diameter (150 nm) and the estimated effective NC diameter (180 nm) are much larger than what could be realized using state-of-the-art MTJ lithography.
We see no fundamental reason against fabricating NCs down to 30 nm, which would then translate to wavelenghts down to 15 nm and SW frequencies well beyond 300 GHz. The use of higher order propagating SW modes might therefore be the preferred route towards ultra-high frequency STNOs.

\begin{methods}
\subsection{\label{sec:level2}MTJ multilayer.}
The magnetron sputtered MTJ stack contains two CoFeB/CoFe layers sandwiching a MgO tunneling barrier with a resistance-area (RA) product of 1.5 $\Omega\ \mu m^2$ \cite{Silva2013ieeetm,Costa2015ieetm,Martins2017}. The top CoFeB/CoFe bilayer acts as the free layer (FL) and the bottom one as the reference layer (RL). A pinned layer (PL) is made of CoFe which is separated from the RL by a Ru layer. An antiferromagnetic PtMn layer is located right below PL. The complete layer sequence is: 
Ta(3)/CuN(30)/Ta(5)/PtMn(20)/CoFe$_{30}$(2)/Ru(0.85)/\\CoFe$_{40}$B$_{20}$(2)/CoFe$_{30}$(0.5)/MgO/CoFe$_{30}$(0.5)/\\CoFe$_{40}$B$_{20}$(1.5)/Ta(3)/Ru(7), with thicknesses in nm.

\subsection{\label{sec:level2}Nanocontact fabrication and static characterization.}
After stack deposition, 16 $\mu$m $\times$ 8 $\mu$m mesas are defined using photolithography. To make the hybrid NC structure, electron beam lithography (EBL) with a negative tone resist is used to define nanocontacts with a nominal diameter of 150 nm.  The scanning electron microscopy (SEM) image of a sample after the development of the exposed resist is shown in Fig. 1(b). The negative tone resist is used as an etching mask in the ion beam etching (IBE) process. Etching of the cap layers in IBE is carefully monitored by in-situ secondary ion mass spectroscopy to prevent any damage to the layers underneath the cap. After this step a structure similar to that shown in Fig 1(a) is realised. Following the etching process, 30 nm of $\mathrm{SiO}_{2}$ to provide electrical insulation between the cap and top contact. The remaining negative tone resist acts as a lift-off layer this time. The devices are left in hot bath of resist remover combined with a high-energy ultrasonic machine for a successful lift-off.  Fig. 1(c) shows the final MTJ mesa after $\mathrm{SiO}_{2}$ lift off. G-S-G labels in this figure, represent the ground-signal-ground contacts in the mesa. In order to provide electrical access to the devices, top contact is defined using photolithography as can be seen in Fig. 1(d).

The static magnetic states throughout key points of the reversal are highlighted as insets in Fig.~1(c) \& (d). Decreasing the field from a fully saturated state ($1$) allows the reference layer (RL) to gradually rotate to be anti-parallel ($2$) with the pinned layer (PL) due to the strong antiferromagnetic coupling (AFC). In going from state $2 \rightarrow 3$, the free layer (FL) switches rapidly in a relatively small field and once again becomes parallel to the RL; hence a minimum $R$ is restored. Note that the FL minor loop is shifted towards positive fields in both Fig.~1(c) and Fig.~1(d), indicating some weak ferromagnetic coupling to the RL. Upon further decreasing the field, the magnetic state moves from $3 \rightarrow 4$, as the PL, working against the strong AFC and weaker exchange bias, slowly switches to be parallel to the RL. In Fig. 1(d), we find small increases in $R$ when moving from states $3 \rightarrow 4$ and $4 \rightarrow 5$. These can be attributed to minor scissoring of the RL and PL layers due to their strong AFC. As one goes from state $5 \rightarrow 6 \rightarrow 1$, the FL switches to align with the applied field followed by the RL switching at high field. 

\subsection{Ferromagnetic resonance (FMR) measurements.}

The magnetodynamic properties of the free layer (CoFeB) are determined using an un-patterned thin films stack. The inset of Fig.~1(d) shows the extracted FL resonance field from broadband FMR measurements (blue squares), fitted with the standard Kittel equation (red line). From the fit we extract the values of the gyromagnetic ratio, $\gamma$/2$\pi$\,=\,29.7 GHz/T, and saturation magnetization, $\mu_{0}M_{s}$\,=\,1.41 T. Subsequent microwave measurement are performed such that the in-plane component of the field lies along the EA of the MTJ stack. We also study the temperature dependence of the magnetodynamcis at low temperature using a NanOsc Instrument CryoFMR system. The low temperature measurements are performed between 10 K-340 K. At each temperature, the FMR response was measured at several frequencies over the range 4--16 GHz where an external magnetic field is applied in the film plane. At each frequency, the resonance field ($H_{res}$) is extracted by fitting the FMR to a Lorentzian function. We extracted the effective magnetization ($M_{eff}$) of CoFeB thin films by fitting the dispersion relation (frequency vs field) to the Kittel equation $f = \frac{\gamma \mu_0}{2\pi} \sqrt{H(H+M_{eff})}$, where $\frac{\gamma}{2\pi}$ is the gyromagnetic ratio. We fit the variation of $ M_{eff}$ with the temperature to a Bloch-law to extract $M_{eff}$ at higher temperatures (T $>$ 340 K).

\subsection{Microwave measurements.}
All measurement were performed at room temperature. In-plane magnetization hysteresis loops of the blanket MTJ multilayer film stacks
were measured using an alternating gradient magnetometer (AGM).  The magnetoresistance (MR) was measured using a custom-built four-point probe station.
The magnetodynamic properties of the unpatterned free layer were determined from using a NanOsc Instruments PhaseFMR spectrometer. .  

Microwave measurement were performed using a probe station with a permanent magnet Halbach array producing a uniform and rotatable out-of-plane field with a fixed magnitude $\mu_{0}H=0.965$ T. 
A direct electric current, $\mathrm{I}_{\mathrm{dc}}$, was applied to the devices through a bias tee and the resulting magnetodynamic response was first amplified using a low-noise amplifier and then measured electrically using a 40 GHz spectrum analyzer. Microwave measurements at higher fields were performed using another custom-built setup capable of providing a uniform magnetic field of up to $\mu_{0}\mathrm{H}=1.8$ T.  
$10\hspace{0.15cm}\mu \mathrm{s}$-long single-shot time domain measurements were performed using a real-time oscilloscope with a sampling rate of 80 GS/s and bandwidth of 30 GHz.

\subsection{Bullet frequency and PSW spectrum calculation.}
The angular dependence of the nonlinear frequency coefficient, \textit{N}, is calculated from the following expression\cite{gerhart2007prb}:
\begin{equation}
\textit{N} =\frac{ f_{H} f_{M}}{f_{FMR}} \left(\frac{3 f_{H}^2 sin\theta_{int}^2}{f_{FMR}^2}-1\right), 
\end{equation}
where $f_{FMR}$ is the FMR frequency, $f_{H}$ = $\frac{\gamma}{2\pi} \mu_0 H_{int}$, $f_{M}$ = $\frac{\gamma}{2\pi} \mu_0 \textit{M}_{s}$, and finally  $H_{int}$ and $\theta_{int}$ are the \emph{internal} magnetic field magnitude and out-of-plane angle, respectively.
 ${H}_{int}$ and $\theta_{int}$ are extracted using a magnetostatic approximation:
\begin{equation}
\label{eq-internal}
\textit{H}_{ex}\hspace{0.15cm}cos\theta_{ex} = \textit{H}_{int}\hspace{0.15cm}cos\theta_{int},
\end{equation}
\begin{equation}
\textit{H}_{ex}\hspace{0.15cm}sin\theta_{ex} = (H_{int} + M_{s})\hspace{0.15cm}sin\theta_{int}.
\end{equation}

The frequency of the Slavin-Tiberkevich bullet mode is calculated from\cite{slavin2005prl}:
\begin{equation}
f_{SWB} = f_{FMR} + NB_{0}^2,
\end{equation} 
where $f_{SWB}$ is the bullet angular frequency, and $B_{0}$ is the characteristic spin wave amplitude. The calculated $f_{SWB}$ quantitatively describes the measured field dependence by setting $B_{0}$=0.46, providing further evidence that this mode is, in fact, a solitonic bullet. The value of  $B_{0}$ is calculated according to Tyberkevych \emph{et. al.}\cite{slavin2005prl} and reaches  $B_{0}$=0.46, which is the upper limit of the theory. 
The spectrum of the propagating spin waves in the linear limit is defined as
\begin{equation}
f_{0}(k) = \frac{\gamma}{2\pi}\sqrt{(H_{int}+D k^2)(H_{int}+M_{eff}\cos \theta_{int}+D k^2)}, 
\end{equation} 
where $D=2A_{ex}/M_{eff}$ is the dispersion coefficient, and $A_{ex}$ is the exchange stiffness constant.
\end{methods}

\begin{addendum}
 \item[Acknowledgements] This work was supported by the European Commission FP7-ICT-2011-contract No. 317950 ``MOSAIC''. It was also supported by the European Research Council (ERC) under the European Community’s Seventh Framework Programme (FP/2007-2013)/ERC Grant 307144 ``MUSTANG''. Support from the Swedish Research Council (VR), the Swedish Foundation for Strategic Research (SSF), the G\"oran Gustafsson Foundation, and the Knut and Alice Wallenberg Foundation is gratefully acknowledged.
 \item[Author contributions] A.H. fabricated the devices and performed all the microwave measurements. R.K. performed analytic calculations. M.H. performed the FMR measurements. R. F. and P. P. F. provided the MTJ stack. R.K.D. co-supervised the project. J.\AA. initiated and supervised the project. All authors contributed to the data analysis and co-wrote the manuscript.
 \item[Competing Interests] The authors declare that they have no competing financial interests.
 \item[Correspondence] Correspondence and requests for materials should be addressed to

J. \AA{}~(email: johan.akerman@physics.gu.se).
\end{addendum}



\bibliographystyle{naturemag}

\end{document}